\documentclass[iop,twocolumn,apjl,letterpaper]{emulateapj}

\usepackage{lipsum}
\usepackage{graphicx}
\usepackage{amsmath}
\usepackage{rotating}
\begin{document}
\title{Stellar Winds and Dust Avalanches in the AU Mic Debris Disk}

\author{Eugene Chiang\altaffilmark{1,2,4} \& Jeffrey Fung\altaffilmark{1,3,4}}
\altaffiltext{1}{Department of Astronomy, University of California at Berkeley, Campbell Hall, Berkeley, CA 94720-3411}
\altaffiltext{2}{Department of Earth and Planetary Science, University of California at Berkeley, McCone Hall, Berkeley, CA 94720-3411}
\altaffiltext{3}{NASA Sagan Fellow}
\altaffiltext{4}{The author list is in alphabetical order; all authors
contributed equally to the intellectual content
and to the work load.}

\email{email: echiang@astro.berkeley.edu, jeffrey.fung@berkeley.edu}

\begin{abstract}
We explain the fast-moving, ripple-like features in the
edge-on debris disk orbiting the young M dwarf AU Mic.
The bright features are clouds of sub-micron dust 
repelled by the host star's wind.
The clouds are produced by avalanches: 
radial outflows of dust that gain exponentially more
mass as they shatter background disk particles in collisional
chain reactions. The avalanches are triggered from 
a region a few AU across---the ``avalanche zone''---located on
AU Mic's primary ``birth'' ring, at a true distance of $\sim$35 AU
from the star but at a projected distance more than a factor of 10 smaller:
the avalanche zone sits directly along the line of sight to the
star, on the side of the ring nearest Earth, launching clouds
that disk rotation sends wholly to the southeast, as observed.
The avalanche zone marks where the primary ring intersects a secondary ring
of debris left by the catastrophic disruption of a progenitor
up to Varuna in size,
less than tens of thousands of years ago. 
Only where the rings intersect
are particle collisions sufficiently violent to spawn the sub-micron
dust needed to seed the avalanches. We show that this picture
works quantitatively, reproducing the masses, sizes, and
velocities of the observed
escaping clouds. The Lorentz force exerted 
by the wind's magnetic field, whose polarity reverses periodically
according to the stellar magnetic cycle, promises to explain the
observed vertical undulations.
The timescale between avalanches, about 10 yr, might be set 
by time variability of the wind mass-loss rate 
or, more speculatively, by some self-regulating limit cycle.
\end{abstract}

\section{INTRODUCTION}
\label{intro}

Dust in debris disks originates from collisional cascades.
The largest bodies, comprising the top of the cascade, have lifetimes
against collisional disruption equal to the system age, 
tens of Myrs or longer.
They grind down into particles micron-sized or smaller 
at the bottom of the cascade. These tiny particulates
are blown out of the system,
typically by stellar radiation
pressure, on orbital timescales of tens to thousands of
years. For a general review of debris disks,
see \citet{matthews14}.

Quasi-steady cascades, in which the rate of mass erosion
is constant from top to bottom (e.g., \citealt{dohnanyi69};
\citealt{pan05}; \citealt{wyatt11}), offer a ready framework for
modeling debris disks 
on the longest of evolutionary timescales
\citep[e.g.,][]{wyatt08,lohne08,gaspar13}. 
At the same time, there are increasingly many observations
demanding that we resolve our theories more finely,
both in space and time, and accommodate more stochastic phenomena. 
Mid-infrared excesses that are unusually strong given the  
Gyr ages of their host stars are thought to
signal recent catastrophic collisions or sudden comet showers 
\citep[e.g.][]{song05,beichman05,wyatt05,weinberger11,kennedy13}. Fast time variability in the infrared, on timescales of months
to years, has been interpreted
as tracing the immediate aftermath of a giant plume-inducing impact
(\citealt{meng14}; see also, e.g.,
\citealt{kenyon05}; \citealt{melis12}; \citealt{kral15}).

Of all the short-timescale phenomena reported for debris disks,
perhaps the most surprising and least understood is
the discovery by the SPHERE (Spectro-Polarimetric High-contrast Exoplanet
REsearch) team of fast-moving
features in the AU Mic edge-on debris disk \citep{boccaletti15}.
The features appear as intensity variations
at projected stellocentric separations of
$\sim$10--50 AU, and are seen only on the southeast ansa of the disk.
They travel away from the star at projected speeds
comparable to---and for the most distant features
exceeding by a factor of $\sim$2---the system escape velocity.
The features also appear undulatory;
the ones closest to the star are elevated by an AU or so
above the disk midplane
(see also \citealt{schneider14} and
\citealt{wang15}).

Taken at face value, the faster-than-escape velocities,
and the trend of increasing velocity with increasing
stellar separation, suggest that the brightest features
are coherent ``clouds'' of dust accelerated radially away
from the star by a force
stronger than the star's gravity by a factor 
on the order of 10 \citep{sezestre17}.
We will adopt this simple interpretation. 
The fact that the clouds are seen to only one side
of the star, together with the recognition
that most of the mass in the underlying
disk is concentrated in a ``birth ring'' of radius $\sim$35 AU
(\citealt{augereau06}; \citealt{strubbe06}),
suggests that the clouds are launched from the birth ring---from the
side of the ring nearest the observer,
so as to appear bright in forward-scattered
starlight---at a special azimuthal location lying
directly along the observer's line
of sight to the star. Then the clouds simply inherit the
orbital motion of the birth ring, whose near side must rotate from
the northwest to the southeast to send the clouds to the southeast.

The host star's wind 
can generate, for sufficiently small grains,
the required radially outward force,
of magnitude
$\beta_{\rm w}$ relative to stellar gravity:
\begin{align}
\beta_{\rm w} &\equiv \frac{3 \dot{M}_\ast v_{\rm wind}}{16 \pi G M_\ast \rho_{\rm p} s} \nonumber \\
&\sim 4 \left( \frac{\dot{M}_\ast}{10^3 \dot{M}_\odot} \right) \left( \frac{v_{\rm wind}}{400 \, {\rm km/s}} \right) \left( \frac{0.1 \, \mu{\rm m}}{s} \right) \label{eq:beta}
\end{align}
where $G$ is the gravitational constant; $M_\ast = 0.6 M_\odot$ is the
stellar mass \citep{boccaletti15};
grains are assumed spherical with internal bulk density
$\rho_{\rm p} \sim 1$ g/cm$^3$ and radius $s$; and $v_{\rm wind}$
and $\dot{M}_\ast$ are the stellar wind's speed and mass-loss rate,
with the latter scaled 
to the solar mass loss rate
$\dot{M}_\odot = 2 \times 10^{-14}M_\odot$/yr \citep{cohen11}.
Grain porosity in AU Mic \citep{graham07,shen09}
may boost $\beta_{\rm w}$ by an extra factor 
on the order of 2.
Augereau \& Beust (\citeyear{augereau06}; see also \citealt{schuppler15}) 
lay out the many reasons why the mass loss rate from this young,
active M dwarf is orders of magnitude larger
than the solar mass loss rate. See in particular their
Figure 11, which attests that $\beta_{\rm w}$ can be
as large as $\sim$40 when AU Mic flares.\footnote{\citet{strubbe06}
argue against high mass loss rates in AU Mic, relying instead on
stellar radiation pressure to blow
out grains. However, they idealized the grain cross section
to radiation pressure as geometric; this is an overestimate
given real-life optical constants
(e.g., Figure 1 of \citealt{schuppler15}).}

We propose here an explanation for the escaping clouds in AU Mic.
We posit that 
they are the outcome of dust avalanches:
exponential rises
in dust production caused by small grains moving on unbound
trajectories and shattering bound disk material in their path
\citep{artymowicz97, grigorieva07}.
In an avalanche, sub-micron grains accelerated to high
radial speeds (in this case by the powerful stellar wind) collide with
larger parent bodies in the birth ring to create still more
sub-micron grains; these collisional
progeny are themselves brought up to high speed,
leading to a collisional chain reaction and exponential
amplification of the escaping dust column. We propose that
each of the bright, fast-moving features observed by
\citet{boccaletti15} results from an avalanche,
launched from a small
region (a few AU in size)
in the birth ring
lying directly along the line of the sight to the star (see above).
Only in this
localized region---what we call the ``avalanche zone''---are
sub-micron grains
produced that can seed the avalanche. In our model, the avalanche zone
marks where the birth ring is intersected by another structure: a
secondary ring, much less massive than the primary, substantially inclined
and/or eccentric, and composed of debris from the
catastrophic disruption of a planetesimal. Avalanches
are triggered at the intersection point of these two rings,
where collisions are especially violent.

Most of the rest of this paper is devoted to reproducing
the sizes, masses, and velocities
of individual escaping clouds using avalanches.
Regarding what sets the periodicity of the avalanches, we have less
to say. To match the observations, the avalanche period
must be the time between cloud ejections, i.e., 
the projected separation between clouds divided by their
projected velocity. An approximate, characteristic value for
the cycle period is
\begin{align}
t_{\rm cycle} &\sim \frac{10 \, {\rm AU}}{5 \,{\rm km}/{\rm s}} \nonumber \\
&\sim 10 \,{\rm yr} \,. \label{eq:tcycle}
\end{align}
We can imagine two mechanisms that can set this period.
The first is time variability in the stellar mass loss rate.
This proposal is admittedly somewhat ad hoc.
Although AU Mic flares dramatically at ultraviolet
and X-ray wavelengths
(e.g., \citealt{robinson01}; \citealt{augereau06} and references
therein), these bursts of high-energy radiation last mere minutes,
whereas equation (\ref{eq:tcycle}) indicates
that we are interested in modulating stellar activity on timescales
of years. It is not clear whether the
similarity between our required value for $t_{\rm cycle}$
and the Sun's 11-year period for magnetic field reversals 
should be regarded as encouraging or irrelevant. An argument in favor
of the latter is that the solar wind mass-loss rate 
$\dot{M}_\odot$ betrays no correlation with solar magnetic cycle
\citep{cohen11}. On the other hand, the rate
of solar coronal mass ejections (CMEs) increases by an order of magnitude
from solar minimum (when the CME rate is 0.5/day) to solar maximum
(when the rate is 6/day; \citealt{gopalswamy03}).
Magnetic activity cycles 
are just beginning to be photometrically detected
for main-sequence stars {\it en masse}
(\citealt{reinhold17}; there are rumors of a weak correlation 
between magnetic cycle period and stellar
rotation period). For pre-main-sequence
stars like AU Mic, there seem to be no useful data on magnetic
cycles or on the time evolution of their mass loss rates, 
although the
situation may be changing with long-term monitoring by 
the Las Cumbres Observatory Global Telescope network
(LCOGT; \citealt{brown13}).

An alternative idea is that the stellar wind
blows strongly (as is demanded by the observations
which point to large $\beta_{\rm w}$)
but steadily on yr-to-decade timescales,
and that the avalanches undergo some kind of
self-regulating limit cycle.
It takes time for avalanches to clear and for enough material
to re-seed them; this time could be identified with $t_{\rm cycle}$.
We will briefly attempt to make this identification at the end
of this paper, after quantifying some of the avalanche dynamics
in \S\ref{model}. 
We do not, however, provide an actual limit-cycle model;
in particular, we do not answer the key question of why
avalanches would not simply unfold in a steady fashion if the
stellar wind were steady.

Since the idea of a self-regulating
cycle is nothing more than a speculation
at this point, we will assume throughout this paper
that the avalanche period $t_{\rm cycle}$ is set by the 
period of a variable stellar mass-loss rate.
Fortunately, many of the remaining elements of our
proposal do not depend on this assumption. We flesh our model out 
in \S\ref{model}. A summary, including some predictions 
and a recapitulation of unresolved issues, is given in \S\ref{sum}.
In this first cut at a theory,
we aim throughout for order-of-magnitude accuracy only.

\section{MODEL}
\label{model}

We present an order-of-magnitude understanding of the
escaping clouds in the AU Mic system. 
We set the stage in \S\ref{cloud_mass} by estimating
individual cloud masses and mass ejection rates: these
are the observables which any theory must explain.
The heart of our paper is in \S\ref{avalanche},
where we sketch our picture of dust avalanches,
detailing quantitatively the creation of an avalanche zone
from an ancient catastrophic collision; the current properties
of the zone; and how the zone can give rise to the observed
escaping clouds. In \S\ref{mag}
we briefly consider magnetic levitation of grains in an attempt
to explain the observed vertical displacements of the clouds.
To illustrate 
and provide a proof of concept of our ideas,
we offer a numerical simulation in \S\ref{sim}.

\subsection{Cloud Mass and Mass Loss Rate}
\label{cloud_mass}

We estimate the mass of a given cloud 
using feature``B'' \citep{boccaletti15} as a fiducial.
The V-band surface brightness of cloud B is comparable
to that of the local disk, about $B \sim 16 \,{\rm mag}/{\rm arcsec}^2 
\sim 0.5 \, {\rm erg}/({\rm cm}^2 \,{\rm s}\, {\rm sr})$ 
(\citealt{krist05}; \citealt{schneider14}).
We take the line-of-sight column density of grains in the cloud
to be $N$; the scattering cross section per grain to be
$Q \pi s^2$; the relative
power scattered per grain per steradian to be $P$
(normalized such that its integral over all solid angle equals
unity); and $a$ to be the true (not projected)
distance between the cloud and the
star of luminosity $L_\ast \sim 0.1 L_{\odot}$. Then
\begin{equation}\label{eq:B}
B \sim \frac{L_\ast}{4\pi a^2} N P Q \pi s^2 \,.
\end{equation} 
The cloud mass is
\begin{equation} \label{eq:M}
M_{\rm cloud} \sim \frac{4}{3} \pi \rho_{\rm p} s^3 N A
\end{equation}
where $A \sim 4 \,{\rm AU}$ (length) $\times$ $2 \,{\rm AU}$ (height) 
is the projected area of the cloud.
Combining (\ref{eq:B}) and (\ref{eq:M}), we have
\begin{align}
M_{\rm cloud} &\sim \frac{16 \pi \rho_{\rm p} B A s a^2}{3QPL_\ast} \nonumber \\
  &\sim 4 \times 10^{-7} M_\oplus \left( \frac{s}{0.1 \,\mu{\rm m}} \right) \left( \frac{0.3}{Q}\right) \times \nonumber \\ 
  & \left(\frac{2/4\pi}{P} \right) \left( \frac{a}{35 \,{\rm AU}} \right)^2 \label{eq:m2}
\end{align}
where we have placed the cloud near the birth ring at
\begin{equation}
a \sim 35 \, {\rm AU}
\end{equation}
and allowed for some forward scattering of starlight
by assigning $P$ to be the isotropic scattering value $1/4\pi$
multiplied by a factor of 2. We adopt throughout this paper
a nominal cloud grain size of 
\begin{equation}
s \sim 0.1 \, \mu{\rm m} \,,
\end{equation}
small enough
for grains to enjoy high acceleration (equation \ref{eq:beta})
but still large enough to scatter starlight with reasonable efficiency.

The implied mass loss rate from the debris disk---from
clouds only\footnote{Equation (\ref{eq:mdot}) accounts
only for the mass ejected in clouds (overdensities). There is also the mass
ejected in inter-cloud regions, which we will account for
in \S\ref{check}. \label{foot:account}}---is
\begin{equation} \label{eq:mdot}
\dot{M}_{\rm cloud} \sim \frac{M_{\rm cloud}}{t_{\rm cycle}} \sim 4 \times 10^{-8} M_\oplus / {\rm yr} \,.
\end{equation}
This is a large rate: it would take only $\sim$0.25 Myr,
or $\sim$1\% of the stellar age of
$t_{\rm age} \sim 23$ Myr \citep{mamajek14}, 
to drain the disk of $\sim$$0.01 M_\oplus$,
the total disk mass inferred from millimeter-wave
observations \citep{matthews15}.
This calculation suggests that the episodic ejections
we are currently observing have not persisted
for the system age, but reflect instead a transient phase. 
We will provide additional support for
this idea in \S\ref{secring} and \S\ref{check}.

\subsection{Avalanche Dynamics}
\label{avalanche}

\subsubsection{The Azimuth of the Avalanche Zone, Where Clouds are Launched}
\label{azimuth}
We consider a $\beta_{\rm w}$-avalanche composed of 0.1 $\mu$m
grains accelerated radially across the birth ring.
The avalanche occurs in a restricted region on the birth ring---the 
``avalanche zone''---that is fixed in inertial space. In other words,
the azimuth of the zone does not rotate at Keplerian speed (the next
section \ref{intersect} explains why).

Because ($i$) the birth ring has a radius of $\sim$35 AU while the 
cloud closest to the star (``A'') is located at a projected
separation of $\sim$8 AU, and ($ii$) all moving features are located
on the southeast ansa and travel further southeast,
we position the azimuth of the avalanche zone 
practically directly along the line of sight to the
star, at a projected stellar separation $\ll$ 8 AU,
so that clouds launched from there can be delivered by disk
rotation to the southeast.
We also locate the avalanche zone on the half of the ring
nearest the observer so that the dust clouds it produces
appear bright in forward-scattered light.

\subsubsection{The Avalanche Zone Lies Where a Secondary Debris Ring Intersects the Birth Ring; Violent Collisions Here Produce Avalanche Seeds} \label{intersect}

\begin{figure}[]
\includegraphics[width=0.99\columnwidth]{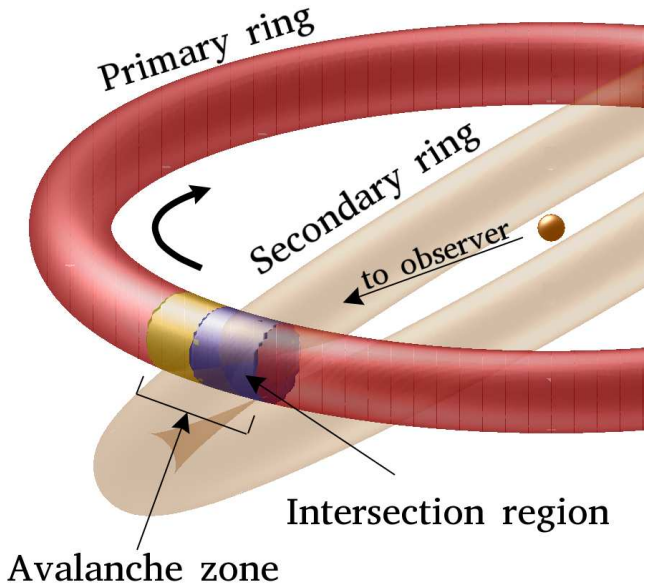}
\caption{Schematic of our model for AU Mic. The primary
(a.k.a.~``birth'') ring (red) is pierced by a secondary ring 
(translucent) at the ``intersection region'' (blue)
where primary and secondary ring
particles smash into each other at velocities fast enough to
generate small 0.1-$\mu$m grains. These grains are accelerated radially
by the stellar wind to km/s speeds, driving collisional chain reactions
with primary ring particles in the ``avalanche zone'' (blue + yellow).
The avalanche zone includes the intersection region and can extend
slightly further if seeds have time to travel azimuthally 
before the 
stellar wind 
flushes them out of the system altogether
(the direction of primary ring rotation is indicated by the curved arrow).
The fast-moving bright features observed by \citet{boccaletti15}
are clouds of dust launched from the avalanche zone.
The ray from the star to the intersection region/avalanche zone
points to Earth. The secondary ring is composed of debris from
the catastrophic disruption of a planetesimal less than
a few hundred km in size. Although we have drawn the secondary ring
inclined relative to the primary, it need not be;
an eccentric secondary ring is also possible, so long as it
creates an intersection region spanning just a few AU.
}
\label{fig:model}
\end{figure}

In our model, the avalanche zone
is rooted where the
birth ring---hereafter the ``primary''---intersects 
a much less massive ``secondary'' ring composed of debris from
a catastrophically disrupted body (we will place
an upper bound on its mass in \S\ref{secring}). 
The node where the rings intersect is stationary in inertial space
(aside from an insignificant precession). 
A similar set-up
was considered in generic terms by \citet{jackson14}; see also the ``static''
case of \citet{sezestre17} (neither of these studies
considered avalanches, and the latter focused on matching
the velocity profile of AU Mic's escaping clouds, not their sizes and
masses). 
See Figure \ref{fig:model} for a big-picture schematic.

We imagine the secondary ring to have a
semimajor axis and therefore an
orbital period comparable to that of the primary,
\begin{equation}
t_{\rm orb} \sim 300 \, {\rm yr}\,,
\end{equation}
and to be substantially inclined 
relative to the primary, reflecting the aftermath of
a collision between a projectile in the primary ring
and a target (the progenitor of the secondary ring) 
that once moved on an orbit inclined to the primary by $\sim$1 rad.
Alternatively, instead of a large mutual inclination,
we could just as well posit
a large eccentricity for the secondary progenitor.
The secondary ring could then be nearly co-planar with the primary,
and be so eccentric that it intersects the primary at the same
special azimuthal location of the avalanche zone.
The only real requirement on the relative ring geometry is that 
this ``intersection region'' be just a few AU large,
in order to match the observed sizes of the escaping clouds
(we elaborate on these considerations of size
in the next section \ref{size}).\footnote{We are
positing only one intersection region, but there can be 
either one or two
if the rings are coplanar and eccentric, or two if
the rings are circular and inclined and have identical radii.
There is no particular reason to think there are two intersection regions
based on the \citet{boccaletti15} observations, which indicate
only a single launch site for grains, but we suppose it is possible
that clouds launched from a second intersection region, on the side of
the primary ring farther from the observer, could
go undetected in back-scattered light.}
We note that the asteroid and Kuiper belts, 
which are solar system analogues of debris disks, 
contain bodies commonly moving on highly inclined 
and eccentric orbits.
Given either a large inclination or large eccentricity for the
secondary ring, the relative velocities
between secondary and primary
particles in the intersection region are large:
\begin{equation} \label{eq:violence}
v_{\rm sec,pri} \sim v_{\rm K}/2 \sim 2 \, {\rm km/s} \,,
\end{equation}
i.e., within factors of a few of the local Keplerian
velocity
\begin{equation}
v_{\rm K} \sim 4 \, {\rm km/s} \,.
\end{equation}

The intersection region 
is where 0.1-$\mu$m ``seeds''
for the avalanche are generated from catastrophic collisions
between primary and secondary ring
particles. The large relative velocities
in the intersection region, which approach if not
exceed the elastic
wave speed in solid rock, readily lead to shattering of 
primary and secondary ring particles 
down to the small, $\sim$0.1-$\mu$m sizes suitable for strong
radial acceleration by the stellar wind (equation \ref{eq:beta}).
We quantify these statements further in \S\ref{seedmass}.

Outside the intersection region, in the rest of the primary
ring, particle relative
velocities are too low to generate sub-micron seeds.
In the bulk of the primary, which is composed of bound and
more nearly micron-sized particles (those dominating the primary's
optical depth), relative velocities are of order
\begin{equation}
v_{\rm pri,pri} \sim 100 \, {\rm m/s}
\end{equation}
as judged from the observed
vertical thickness of the ring \citep{strubbe06}.
Collisions at such velocities 
will chip and erode, but do not lead to catastrophic
disruption, as they correspond to specific kinetic
energies
\begin{equation}
\frac{1}{2} v_{\rm pri,pri}^2 \sim 5 \times 10^7 \, {\rm erg/g}
\end{equation}
that fall short of 
\begin{equation}
S^\ast \sim 2 \times 10^8\, {\rm erg/g} 
\end{equation}
the threshold for catastrophic disruption
of micron-sized, relatively flaw-free targets
(\citealt{grigorieva07}; \citealt{tielens94}).
Thus in non-intersection regions, collisional cascades
are not expected to proceed past particle sizes
for which $\beta_{\rm w} \sim 0.5$,
the minimum threshold for unbinding particle orbits.
By contrast, the violence of collisions
in the intersection region
(equation \ref{eq:violence}) 
permits the creation of especially small grains attaining
$\beta_{\rm w} \gg 1$.\footnote{At the risk of
belaboring the point,
an analogy would be with  
high-energy collisions in a particle accelerator;
the greater the energy of the collision,
the smaller the constituent particles that are unleashed.
See also
\citet{leinhardt12}.}

\subsubsection{Sizes of the Intersection Region 
and \\ of the Avalanche Zone} \label{size}
By definition, the avalanche zone comprises all regions
where $\sim$0.1-$\mu$m grains (accelerating projectiles)
and primary ring particles (field targets that shatter into more  projectiles)
co-exist. The avalanche zone includes the intersection region
where $\sim$0.1-$\mu$m
seeds are created from collisions between primary
and secondary ring particles. Avalanches can also extend
beyond the intersection region because seeds can travel azimuthally 
(at the Keplerian speeds which they inherit at birth from their
primary parents), out of the
intersection region into the rest of the primary ring.
Thus the characteristic dimensions of the avalanche
zone and of the intersection region obey (see Figure \ref{fig:model}):
\begin{equation} \label{eq:lowerbound}
\Delta l_{\rm avalanche} > \Delta l_{\rm intersect} \,.
\end{equation}

The size of the intersection region scales
with the thickness of the secondary ring (really, a torus);
that thickness, in turn, 
scales with the ejecta 
velocities $v_{\rm ej}$ of
the catastrophic collision that gave birth to the
secondary ring:
\begin{equation} \label{eq:torus}
\Delta l_{\rm sec} \sim \frac{v_{\rm ej}}{v_{\rm K}} \, a \sim 4 \left( \frac{v_{\rm ej}}{400 \, {\rm m/s}} \right) \, {\rm AU} 
\end{equation}
(the thickness scales with the dispersion
of orbital elements of the secondary fragments, and that
dispersion scales with the deviation of fragment orbital velocities
from the progenitor's Keplerian velocity).
Ejecta velocities $v_{\rm ej}$ of $\sim$200--400 m/s
are realistic; see, e.g., modeling of the Haumea
collisional family in the Kuiper belt (\citealt{lykawka12}).
Our nominal estimate for $\Delta l_{\rm sec}$ is
comparable to the radial ($\Delta a$) and
vertical ($H$) thicknesses of the primary ring,
each of which is about 3 AU \citep{strubbe06, augereau06}.
The intersection region between secondary and primary
rings should be about $\Delta l_{\rm sec}$ 
large
(lower if the originating
collision occurred toward the margin of the primary, and
higher if the inclination and/or eccentricity of the secondary ring
are smaller; see section \ref{intersect}):
\begin{equation}
\Delta l_{\rm intersect} \sim \Delta l_{\rm sec} \,.
\label{eq:Deltal}
\end{equation}

An upper limit
on $\Delta l_{\rm avalanche}$ is given by the
distance that seeds travel azimuthally through the primary ring
between avalanches (since a given avalanche 
triggered during the high phase of the stellar wind
flushes the primary of all seeds):
\begin{align} \label{eq:upperbound}
\Delta l_{\rm avalanche} &< v_{\rm K} \,t_{\rm cycle} \nonumber \\
&< 8 \, {\rm AU} \,.
\end{align}
The actual length $\Delta l_{\rm avalanche}$ will be
smaller than this because before seeds have had 
a chance to cover an azimuthal distance of 8 AU,
the stellar wind (during its high phase) will have blown
them radially out of the primary.

Equations (\ref{eq:lowerbound})--(\ref{eq:upperbound})
constrain $\Delta l_{\rm avalanche}$ to be several AU.
This is the right order of magnitude:
\citet{boccaletti15} observe that the bright fast-moving
features are $\Delta l_{\rm cloud} \sim 4$ AU in length.
Computing $\Delta l_{\rm cloud}$ from first principles,
starting from the considerations outlined here,
requires that we fold in the detailed time history of the stellar wind
and of the resultant avalanches---this is what we
do in the numerical model of \S\ref{sim}, where we will find that, because of
strong radial outflows and projection effects,
the intersection region practically single-handedly determines
the observed cloud size
(i.e., $\Delta l_{\rm cloud} \sim \Delta l_{\rm avalanche} \sim \Delta
l_{\rm intersect}$).

Our main takeaway point for this subsection
is that, putting aside the various 
order-unity details, 
a secondary ring of debris
created from a catastrophic collision
has the right thickness, namely a few AU (equation \ref{eq:torus}),
to be relevant for the observations by \citet{boccaletti15}.

\subsubsection{The Secondary Ring: Lifetime and Upper Mass Limit} \label{secring}

The ejecta velocities of the originating collision
place an upper limit on the surface escape velocity
of the progenitor: $v_{\rm ej} \sim 400$ m/s
implies a progenitor radius $\lesssim 400 \, {\rm km}$
or equivalently a progenitor mass
\begin{equation} \label{eq:upper}
M_{\rm sec} \lesssim 10^{-4} M_\oplus 
\end{equation}
where we have assumed a progenitor
bulk density of $\sim$2 g/cm$^3$
(only for equation \ref{eq:upper}; elsewhere, for less
compressed grains, we adopt $\rho_{\rm p} \sim 1$ g/cm$^3$).
A progenitor radius of $\sim$400 km is comparable to those 
of large asteroids (e.g., Vesta)
and Kuiper belt objects (e.g., Varuna).

The secondary ring has a finite lifespan because its particles 
are destroyed by collisions with the primary ring.
Every time a secondary ring particle executes an orbit,
it has a probability of colliding with a primary
particle equal to $\tau_{\rm pri}$, the optical depth 
traversed through the primary ring. That optical depth
is on the order of
\begin{equation} \label{eq:taupri}
\tau_{\rm pri} \sim 0.01
\end{equation}
based on detailed
models derived from scattered light images
\citep{augereau06, strubbe06}.
Then the secondary ring disintegrates on a timescale
\begin{align}
t_{\rm sec} &\sim t_{\rm orb}/\tau_{\rm pri} \nonumber \\
&\sim 3 \times 10^4 \, {\rm yr} \left( \frac{0.01}{\tau_{\rm pri}} \right)\,.
\label{eq:tsec}
\end{align}
By ``disintegrate'' we refer only to those secondary ring particles 
small enough to be shattered by the $\mu$m-sized particles
comprising the bulk of the primary's optical depth. Such 
secondary ring particles are likely to have sizes
smaller than several microns. Despite
their restriction in size, such particles may still carry a fair
fraction of the mass of the secondary progenitor, for two reasons.
The first is that in the immediate aftermath of the progenitor's
destruction, ejecta mass is expected to be distributed equitably
across logarithmic intervals in fragment size. This expectation arises from the ``crushing'' law for catastrophic single collisions
\citep{takasawa11, leinhardt12, kral15}.\footnote{The crushing law
should not be confused with the better known (and often abused)
equilibrium cascade describing the long-term comminution of bodies
as derived by Dohnanyi (\citeyear{dohnanyi69}; see also \citealt{pan05}).
The latter does not distribute mass logarithmically evenly, but concentrates
it in the largest fragments; its use is not appropriate for either
short-timescale, non-equilibrium dynamics, or for particle sizes close to the
blow-out limit \citep{strubbe06}.} Second, as secondary ring bodies
collide with one another and establish an equilibrium cascade, particles near
the bottom of the cascade---i.e., $\mu$m-sized particles on marginally
bound orbits---grow enormously
in population because they spend much of their time at the apastra
of orbits made highly eccentric by the stellar wind, away from destructive collisions in the secondary
ring (see Figure 3 of \citealt{strubbe06}). Thus our estimate of the total
secondary ring mass $M_{\rm sec}$ may not be that much greater than  
the mass in $\mu$m-sized secondary ring particles; we will assume in what
follows that they are within an order of magnitude of one another.

Qualified to refer only to secondary particles small enough
to be susceptible to disruption,
the lifetime of the secondary ring $t_{\rm sec}$
(which notably does not depend
on the secondary ring mass) is some three orders of magnitude
shorter than the system age. This aligns with our
earlier suspicion (\S\ref{cloud_mass}) that the phenomenon
of escaping dust seen today is transient. 
In fact, the avalanches may not even last as long as
$t_{\rm sec}$---see \S\ref{check} for the reason why.

\subsubsection{Seed Mass and Avalanche Mass}\label{seedmass}

Within the intersection region, primary and secondary ring particles
destroy each other to produce 0.1-$\mu$m avalanche seeds at a rate
\begin{equation}
\dot{M}_{\rm seed} \sim \frac{M_{\rm sec}}{t_{\rm sec}} \,.
\label{eq:mseeddot}
\end{equation}
A couple of comments regarding this estimate:
first, although (\ref{eq:mseeddot})
appears superficially to account only for the destruction of
secondary ring particles, it actually accounts for the destruction
of primary ring particles as well, by symmetry
($\dot{M}_{\rm seed} \sim M_{\rm sec} / t_{\rm sec} \propto M_{\rm sec} \tau_{\rm pri} \propto M_{\rm sec} M_{\rm pri}$, where $M_{\rm pri} \sim 0.01 M_\oplus$ is the primary ring mass).
Second, underlying statement (\ref{eq:mseeddot}) is the assumption that
when primary and secondary ring particles shatter each other, they
invest an order-unity fraction of their mass into 0.1-$\mu$m grains.
Here again we appeal to
the crushing law for catastrophic single collisions,
which tends to distribute mass logarithmically uniformly across
particle sizes; we are not appealing to any equilibrium
cascade law like Dohnanyi's (see the discussion in the penultimate
paragraph of section \ref{secring}).

The 0.1-$\mu$m avalanche seeds are accelerated radially outward
by the stellar wind over some fraction of $t_{\rm cycle}$.
If the ``high'' phase of the stellar wind lasts
\begin{equation}
t_{\rm high} \sim t_{\rm cycle}/4 
\label{eq:thigh}
\end{equation}
then the seeds attain a radial velocity
\begin{align}
v_{\beta} &\sim \frac{\beta_{\rm w,high} GM_\ast}{a^2} t_{\rm high} \nonumber \\
&\sim v_{\rm K} \left(\frac{\beta_{\rm w,high}}{20}\right) \left( \frac{t_{\rm high}}{t_{\rm cycle}/4} \right) \,. \label{eq:vbeta}
\end{align}
Other values for $t_{\rm high}$ and $\beta_{\rm w,high}$
are possible; only their product matters in (\ref{eq:vbeta}). (Of course, the product cannot be so high
that the larger bodies comprising the primary and secondary rings
also become unbound.)

The 0.1-$\mu$m seeds slam into more typically
$\mu$m-sized primary parent bodies, creating more 0.1-$\mu$m
grains in an exponentially amplifying avalanche.
By the time the avalanche has propagated
across the radial width of the primary ring, it
has traversed an optical depth $\tau_{\rm pri}$
and acquired a mass
\begin{equation} \label{eq:scary}
M_{\rm avalanche} \sim M_{\rm seed} \exp(\eta \tau_{\rm pri})
\end{equation}
(see, e.g., the order-of-magnitude description of avalanches
by \citealt{grigorieva07}).
Here $\eta$ is the number of fragments produced per
catastrophic collision:
\begin{equation} \label{eq:eta}
\eta \sim \frac{(1/2)m_{\rm proj}v_\beta^2}{S^\ast m_{\rm frag}} \sim \frac{v_{\beta}^2}{2S^\ast}
\end{equation}
where the projectile mass $m_{\rm proj}$
and the individual fragment mass $m_{\rm frag}$
are assumed comparable---both are imagined
to correspond to 0.1-$\mu$m 
grains. 
Inserting (\ref{eq:vbeta}) into
(\ref{eq:eta}) yields
\begin{equation} \label{eq:etanum}
\eta \sim 400 \,.
\end{equation}
An upper limit on $\eta$ can be estimated
by noting that the catastrophic disruption of a 
1-$\mu$m primary parent particle can yield no more than
$\eta_{\rm max} = 1000$ fragments each of size 0.1 $\mu$m.


The seed mass underlying a given cloud is that generated
during the high phase of the stellar wind
(seeds generated
during the low phase give rise to the inter-cloud emission;
see section \ref{check}):
\begin{align}
M_{\rm seed} &\sim \dot{M}_{\rm seed} \times t_{\rm high}  \nonumber \\
&\sim 10^{-8} M_\oplus \left( \frac{M_{\rm sec}}{10^{-4} M_\oplus} \right) \left( \frac{\tau_{\rm pri}}{0.01} \right) \left( \frac{t_{\rm high}}{t_{\rm cycle}/4} \right) \label{eq:mseed} \,.
\end{align}
Putting (\ref{eq:scary}), (\ref{eq:etanum}), and (\ref{eq:mseed}) 
together, we derive a single-avalanche mass of
\begin{align}
M_{\rm avalanche} 
&\sim 5 \times 10^{-7} M_\oplus \left[ \frac{\exp (\eta \tau_{\rm pri})}{50} \right] \left( \frac{M_{\rm sec}}{10^{-4} M_\oplus} \right) \label{eq:mava}
\end{align}
which is a remarkably good match to the observationally
inferred cloud mass $M_{\rm cloud} \sim 4 \times 10^{-7} M_\oplus$
(equation \ref{eq:m2}), considering that we have not
fine-tuned any of the input parameters.

Of course, uncertainties in
$\eta$ and $\tau_{\rm pri}$ will be exponentially amplified
in the avalanche gain factor
$\exp(\eta \tau_{\rm pri})$. Increasing the gain factor would require
that we reduce the secondary ring mass $M_{\rm sec}$
to maintain the agreement between $M_{\rm avalanche}$
and $M_{\rm cloud}$. Thus we can do no better than
re-state our upper bound of $M_{\rm sec} \lesssim 10^{-4} M_\oplus$
(equation \ref{eq:upper}), which in turn implies
that avalanche gain factors $\exp(\eta \tau_{\rm pri}) \gtrsim 50$.

Although equation (\ref{eq:scary}) oversimplifies
the avalanche dynamics (among other errors, it neglects the finite
acceleration times and differing velocities of grains), our conclusions
do not depend on the specific implementation of a simple
exponential to describe avalanches. Stripped to its essentials, our
reasoning can be recapitulated as follows:
the mass of an individual cloud is
$M_{\rm cloud} \sim 4 \times 10^{-7} M_\oplus$ by 
equation (\ref{eq:m2}); generically, the mass unleashed in an
avalanche is $M_{\rm avalanche} \sim M_{\rm seed} \, \times$ Gain,
where Gain $> 1$ need not take the form of a simple exponential;
$M_{\rm seed} \lesssim 10^{-8} M_\oplus$ by equations (\ref{eq:mseed})
and (\ref{eq:upper}); then for $M_{\rm avalanche}$ to match $M_{\rm cloud}$,
we need Gain $\gtrsim 40$. Numerical simulations of avalanches
by Q.~Kral \& P.~Th\'ebault
(personal communication 2017; see also \citealt{grigorieva07}) 
indicate that such gain factors are possible,
even though they are not described by the simplistic exponential
in equation (\ref{eq:scary}).

\subsubsection{Avalanche Propagation Time}
The timescale for the avalanche to propagate radially
across the zone (whose size is
comparable to the radial width of the primary ring; see
\S\ref{size}) is
\begin{align}
t_{\rm rad,esc} &\sim \Delta l_{\rm avalanche} / v_\beta \nonumber \\
&\sim 5 \,{\rm yr} \left( \frac{\Delta l_{\rm avalanche}}{4 \, {\rm AU}} \right) \left( \frac{20}{\beta_{\rm w,high}} \right) \, \label{eq:prop}
\end{align}
which is both shorter than $t_{\rm cycle} \sim 10$ yr,
implying that only
one avalanche occurs per stellar cycle, and also
longer than our assumed acceleration time of
$t_{\rm cycle}/4 \sim 2.5$ yr, as required for consistency.

\subsubsection{Total Mass Budget: Cloud + Inter-Cloud Regions, and Starving the Avalanche} \label{check}
That we can reproduce the observed $M_{\rm cloud}$ (equation \ref{eq:m2})
using our theory for $M_{\rm avalanche}$ (equation \ref{eq:mava})
is encouraging to us. The theory relies on a variety of
estimates, several of which were made {\it a priori},
and it is heartening that the numbers hang together
as well as they do.

Here is another check on our work. Suppose (just for the sake
of making this check; we will see at the end of this subsection
why this supposition is probably not realistic)
that avalanches continue for the entire lifetime of the
secondary ring at their current
pace and magnitude. Then the
total mass lost from the system should equal the mass of the secondary
ring multiplied by the avalanche gain factor:
\begin{equation}
\max M_{\rm total,1} \sim M_{\rm sec} \exp (\eta \tau_{\rm pri}) \sim 0.005 M_\oplus \,.
\end{equation}
(See the caveats regarding our use of $M_{\rm sec}$
in section \ref{secring}.)
We want to check whether this (maximum) mass
matches that inferred more directly from observations of mass loss
(i.e., \citealt{boccaletti15}).
To make this accounting complete, we must include
not only the mass lost in clouds ($\dot{M}_{\rm cloud}$ from equation
\ref{eq:mdot}, multiplied by the secondary ring lifetime
$t_{\rm sec}$ from equation \ref{eq:tsec})
but also the mass lost from
inter-cloud regions. We appeal to our model to account for
the latter.
The inter-cloud regions
represent avalanches from seeds produced during
the ``low'' phase of the stellar wind. 
Because the duration of the low phase is much less than the
orbital time (a.k.a. the system dynamical time),
these seeds are not accelerated much beyond their
Keplerian speeds during the low phase,
and so they stay roughly within the primary
ring during this time. Our numerical
simulation in section \ref{sim} confirms this point---the
low-phase seeds are distributed along streams
that are longer azimuthally than radially.
They do not escape radially
until the stellar wind attains its
high phase, at which time they undergo their own avalanche.
Thus while inter-cloud regions are produced from seeds
generated during the low phase lasting an assumed
$3t_{\rm cycle}/4$, and clouds 
are produced from seeds generated
during the high phase lasting $t_{\rm cycle}/4$,
both sets of seeds are amplified by about the
same avalanche gain factor, because avalanches only occur
when the wind is in its high phase.
Since the seed production rate is constant
(equation \ref{eq:mseeddot}),
the mass-loss rate from inter-cloud regions must be 3$\times$
the mass-loss rate from clouds; the total must be
4$\times$ the latter. Hence
for our second estimate of the total mass lost from
the system, we have
\begin{equation}
\max M_{\rm total,2} \sim 4\dot{M}_{\rm cloud} t_{\rm sec} \sim 0.005 M_\oplus \,.
\end{equation}
The match between $\max M_{\rm total,1}$ and
$\max M_{\rm total,2}$ is better
agreement than we probably deserve.

The maximum total mass lost, $\max M_{\rm total}$,
is still less than what the primary ring contains,
$M_{\rm pri} \sim 0.01 M_\oplus$, but only by about a factor of 2.
The prospect of losing an order-unity fraction of the total disk mass
over a small fraction of the stellar age highlights
the destructive power of avalanches.
But there is good reason to believe that
avalanches will not continue unabated
for the full lifespan of the secondary ring.
If avalanche targets are strictly those at the
bottom of a conventional cascade in the primary ring
($\sim$$\mu$m-sized particles in our simple model, i.e., those
dominating the primary ring's optical depth $\tau_{\rm pri}$), then
avalanches could be ``starved'' if the primary cascade does not
supply such small targets
at a fast enough rate. The primary cascade rate
might only be 
$M_{\rm pri}/t_{\rm age} \sim 0.01 M_{\oplus}/(20 \, {\rm Myr})$,
much lower than the current avalanche mass loss rate
$M_{\rm total}/t_{\rm sec} \sim 0.005
M_{\oplus}/(0.03 \, {\rm Myr})$.
Conceivably avalanches weaken well before $t_{\rm sec}$ elapses
as the population of $\mu$m-sized targets in the primary ring
dwindles. Forecasting the long-term evolution of avalanches
is left for future work.

\subsection{Vertical Oscillations Driven by the Magnetized \\ Stellar Wind}
\label{mag}

The stellar wind, moving with a radial velocity
$v_{\rm wind} \boldsymbol{\hat{r}}$ and
carrying a magnetic field $\boldsymbol{B}$,
exerts a Lorentz force
on dust grains of charge $q$ and velocity $\boldsymbol{v}$:
\begin{equation}
\boldsymbol{F}_{\rm L} = \frac{q}{c} \left[ (\boldsymbol{v} - v_{\rm wind}\boldsymbol{\hat{r}}) \times \boldsymbol{B} \right]
\end{equation}
where $c$ is the speed of light.\footnote{Working in cgs units, in
which electric and magnetic fields have the same units,
and in which the electrical capacitance of a spherical grain
equals its radius $s$.}
At the large stellocentric
distances of interest to us, well outside the wind's Alfv\'{e}n
point, the magnetic field is tightly wrapped
and primarily azimuthal: 
$\boldsymbol{B} \simeq B_{\phi} \boldsymbol{\hat{\phi}}$
\citep{weber67}.
The field strength around AU Mic is unknown, but
we expect it to be larger than in the solar system,
as AU Mic is a rapidly rotating, strongly convective,
active young star---properties that all point to a strong stellar
magnetic field.
For reference,
in the solar system,
$B_\phi \sim 40 \, \mu{\rm G} \, (a/{\rm AU})^{-1}$
\citep[e.g.,][]{schwenn00, weber67}.
Thus we expect that for the AU Mic system
at $a \sim 35$ AU, $B_{\phi} > 1 \,\mu$G.

Given that the field is dominated by its azimuthal component,
and that grain velocities $|\boldsymbol{v}|$
are much smaller than $v_{\rm wind} \sim 400$ km/s,
the Lorentz force is dominated by the
term proportional to
$-v_{\rm wind} \boldsymbol{\hat{r}} \times B_\phi \boldsymbol{\hat{\phi}} = - v_{\rm wind} B_\phi \boldsymbol{\hat{z}}$. 
This term is equivalent to a
vertical electric field
$E_z \boldsymbol{\hat{z}} = - (v_{\rm wind}/c) B_\phi \boldsymbol{\hat{z}}$
(the electric field seen by a grain
when a magnetic field moves past it).
We explore in this subsection how we might use this
vertical electric field to generate the cloud vertical offsets
observed by \citet{boccaletti15}.
We state at the outset that our Lorentz force model
will be found wanting in a few respects when confronted
with observations (\S\ref{sec:unresolved}).

In a simple conception of a stellar magnetic cycle,
the magnetic field varies sinusoidally with period
$t_{\rm mag} = 2\pi/\omega_{\rm mag}$:
\begin{equation}
B_\phi = B_{\phi,0} \cos (\omega_{\rm mag} t) \,.
\end{equation}
The vertical equation of motion for a grain of mass $m$ reads:
\begin{align}
\ddot{z} &= \frac{q E_{z,0}}{m} \cos (\omega_{\rm mag} t) \\
&= \frac{-q v_{\rm wind} B_{\phi,0}}{mc} \cos (\omega_{\rm mag} t) \,.
\end{align}
The solution for the displacement $z$ is oscillatory with a phase
that depends on initial conditions, in particular
the phase in the magnetic cycle at which the grain is born.
A grain born with $z=\dot{z}=0$ at $t=0$ (when the field
$B_\phi$ is strongest with magnitude $B_{\phi,0}$)
is displaced according to
\begin{equation} \label{eq:zplus}
z (t\geq 0) = - \frac{q v_{\rm wind} B_{\phi,0}}{m \omega_{\rm mag}^2 c} [1 - \cos(\omega_{\rm mag} t)] \,;
\end{equation}
it oscillates vertically on one side of the disk, never crossing 
the midplane to the other side.
At the other extreme, a grain born 
with $z=\dot{z}=0$ at $t = \pi/\omega_{\rm mag}$ (when $B_{\phi}$
is strongest with the opposite polarity $-B_{\phi,0}$)
obeys
\begin{equation} \label{eq:zminus}
z (t\geq \pi/\omega_{\rm mag}) = + \frac{q v_{\rm wind} B_{\phi,0}}{m \omega_{\rm mag}^2 c} [1 + \cos(\omega_{\rm mag} t)] 
\end{equation}
and oscillates on the other side. Intermediate
cases cross the midplane.

\subsubsection{Magnitude of Vertical Displacements and Parameter Constraints} \label{pickup}

The maximum vertical displacement is given by
\begin{align}
z_{\rm max} =& \left| \frac{(q / m) B_{\phi,0} v_{\rm wind} t_{\rm mag}^2}{2\pi^2 c} \right| \nonumber \\
=& \,2 \, {\rm AU} \, 
\left( \frac{v_{\rm wind}}{400 \, {\rm km/s}} \right)
\left( \frac{q/m}{5 \times 10^{-8} e/m_{\rm p}} \right) \times \nonumber \\
& \left( \frac{B_{\phi,0}}{30 \,\mu{\rm G}} \right) \left( \frac{t_{\rm mag}}{{\rm yr}} \right)^2 \label{eq:zmax}
\end{align}
where the charge-to-mass ratio $q/m$ is
scaled to the proton value $e/m_{\rm p}$, and 
we have chosen all input parameters to 
yield a vertical offset $z_{\rm max}$ similar to that observed for feature
``A'' by \citet{boccaletti15}. We can relate $q/m$ to the
grain surface potential $\Phi$:
\begin{align}
\frac{q}{m} &= \frac{s\Phi}{4\pi \rho_{\rm p} s^3/3} \nonumber \\
&\sim 5 \times 10^{-8} \frac{e}{m_{\rm p}} \left( \frac{\Phi}{2\, {\rm volt}} \right) \left( \frac{0.1 \,\mu{\rm m}}{s} \right)^2 \,. 
\end{align}
Grains are charged positively by ultraviolet (UV) 
photoelectric emission;
the grain charge equilibrates when the rate at which
photoelectrons are ejected (a process
that becomes less efficient as the grain increases its
positive charge)
balances the rate at which ambient
stellar wind electrons are absorbed.
A surface potential of 2 volts
(SI units; equivalent to 2/300 statvolts in cgs)
would be comparable to potentials for solar
system grains \citep{kimura98}, and would suggest that
when scaling from the solar system to AU Mic,
the increased number of electrons from the stronger stellar
wind nearly balances the heightened UV radiation field.
A first-principles calculation of $q/m$ is reserved
for future work.

Grains are ``picked up'' by the magnetized
wind to reach stellar wind velocities if they are allowed
to complete a magnetic gyration. Since the features
are observed to move with velocities much less than $v_{\rm wind}$,
pick-up must not have occurred
(or at least not fully developed; cf.~\S\ref{sim}).
This constrains
\begin{align}
1 &> \omega_{\rm gyro} t_{\rm mag} \nonumber \\
&> \frac{qB_{\phi,0}}{mc} t_{\rm mag} 
\end{align}
which re-written in terms of (\ref{eq:zmax}) states that
\begin{align}
\frac{2\pi^2 z_{\rm max}}{v_{\rm wind}t_{\rm mag}} &< 1 \nonumber \\
0.5 \left( \frac{z_{\rm max}}{2 \, {\rm AU}} \right) \left( \frac{400 \, {\rm km/s}}{v_{\rm wind}} \right) \left( \frac{{\rm yr}}{t_{\rm mag}} \right) &< 1 \label{eq:nopickup}
\end{align}
implying $t_{\rm mag} > 1$ yr.

\subsubsection{Unresolved Issues with Vertical Deflections}
\label{sec:unresolved}
While our little Lorentz force model predicts an oscillatory vertical motion
that recalls the ``wavy'' structure seen in images of AU Mic
\citep{boccaletti15, schneider14},
the details of this model do not fit the
observations. A key unknown is the period of the
magnetic cycle, $t_{\rm mag}$. One hypothesis
sets $t_{\rm mag} = 2 t_{\rm cycle} = 20$ yr, in the belief that
if the stellar mass-loss rate varies with time---which it would
do with period $t_{\rm cycle}$ by definition---then it would
peak twice per magnetic cycle. In other words, it is imagined
that dust avalanches are launched
at $t = n \pi / \omega_{\rm mag}$ for integer $n$,
when the stellar field is strongest irrespective of polarity.
The hypothesis that $t_{\rm mag} \sim 20$ yr 
runs into the immediate difficulty that
the images of AU Mic, spaced in time
by significant fractions of $t_{\rm mag}$ (they were taken
in July 2010, August 2011,
and August 2014) betray no vertical motion for features A and B;
these clouds appear to float at a practically constant height of
$\sim$2 AU above the midplane at three separate epochs.

Faced with this phasing problem, we might hypothesize instead
that $t_{\rm mag} \sim 1$  yr,
i.e., the observations just happen to catch the clouds
at the same phase. A magnetic cycle period as short as $\sim$1 yr
marginally satisfies (\ref{eq:nopickup}). It 
is not obviously compatible with a stellar-mass
cycle period as long as $\sim$10 yr (but then again,
the stellar-mass loss rate might not vary with this period
in the first place; see \S\ref{intro} and \S\ref{sum}).

Other mysteries include
the observed lack of moving features below the midplane,
and the observation that the vertical offsets 
appear smaller for the most distant clouds.
The expected $1/a$ decay in the azimuthal
magnetic field strength helps to explain this drop off,
but might not be sufficient.

These problems notwithstanding, Lorentz forces still 
appear the most natural way of explaining the observed
vertical undulations. There is no doubt that the stellar
wind is magnetized, emanating as it does from a low-mass
star, just as there is no doubt that sub-micron grains are charged,
bathed as they are in a relatively intense stellar ultraviolet radiation
field. Moreover, the sign of the field must periodically reverse,
because if it
did not, the magnetized wind would eventually pick up grains
and accelerate them to speeds and heights far exceeding
those observed. What we seem to be missing is an understanding of 
the detailed time history of the field, and how it is phased
with the avalanche history.

\subsection{Numerical Simulations}
\label{sim}

We construct numerical simulations to illustrate our ideas,
omitting magnetic fields for simplicity.
We simulate particles that
represent packets of 0.1-$\mu$m grains produced in the 
intersection region and amplified by avalanches. In cylindrical
coordinates centered on the star, the particle equations of motion read:
\begin{align}
\ddot{r} &= -\frac{GM_\ast(1-\beta_{\rm w})}{(r^2+z^2)^{3/2}} r + \frac{l^2}{r^3}  \\
\ddot{z} &= -\frac{GM_\ast(1-\beta_{\rm w})}{(r^2+z^2)^{3/2}} z \,,
\end{align}
where $M_\ast = 0.6 M_\odot$ and $l = r^2 \dot{\phi}$
is the specific angular momentum, conserved 
because there are no
azimuthal forces (torques). 
The force ratio $\beta_{\rm w}$ cycles 
in a step-function manner
between a high value
$\beta_{\rm w,high} = 20 (40)$ lasting
$t_{\rm high} = t_{\rm cycle}/4 = 2.5$ yr, and a low value 
$\beta_{\rm w,low} = \beta_{\rm w,high}/10 = 2 (4)$
lasting $t_{\rm low} = 3t_{\rm cycle}/4 = 7.5$ yr.
All of the numerical parameters
of our simulation are inspired by the various estimates
made in \S\ref{avalanche}.

The simulation particles are initialized as seeds freshly produced in the 
intersection region. They begin their trajectories at $r=r_0=35$ AU,
moving at circular Keplerian speed. Their spatial density 
in the \{$\phi$, $z$\} plane follows a two-dimensional Gaussian
representing the intersection region:
\begin{equation}
I_{\phi,z} \propto \exp \left[ -\frac{r_0^2(\phi-\phi_0)^2 + z^2}{2\sigma^2}\right] \, ,
\end{equation}
where $\phi_0$ points to the observer.
The size of the intersection region is characterized by $\sigma$; we 
choose $\sigma = 1~{\rm AU}$ (the corresponding full width at 2$\sigma$ 
is then 4 AU).

We do not explicitly simulate the avalanche process but model
it as follows. 
Each simulation particle is initialized with a fixed baseline mass.
When that particle experiences $\beta_{\rm w} = \beta_{\rm w,high}$
for the first time at $t = t_1$, its mass (read: light-scattering
cross section)
is subsequently amplified by an avalanche gain factor of 50 at
$t = t_1 + 5$ yr. The delay time of 5 yr represents the
finite propagation time of the avalanche across the primary ring.

We integrate the equations of motion using a second-order leap-frog 
scheme with a fixed time step of $t_{\rm cycle}/500 = 0.02\, {\rm yr}$.
New particles (seeds)
are generated in the intersection region at a constant rate
of $\sim$15000 particles per year. 
When constructing surface brightness maps of the disk viewed edge-on,
we employ a Henyey-Greenstein scattering phase function
with asymmetry parameter $g = 0.25$.

\subsubsection{Simulation Results}
\label{sim_results}

\begin{figure}[]
\includegraphics[width=0.99\columnwidth]{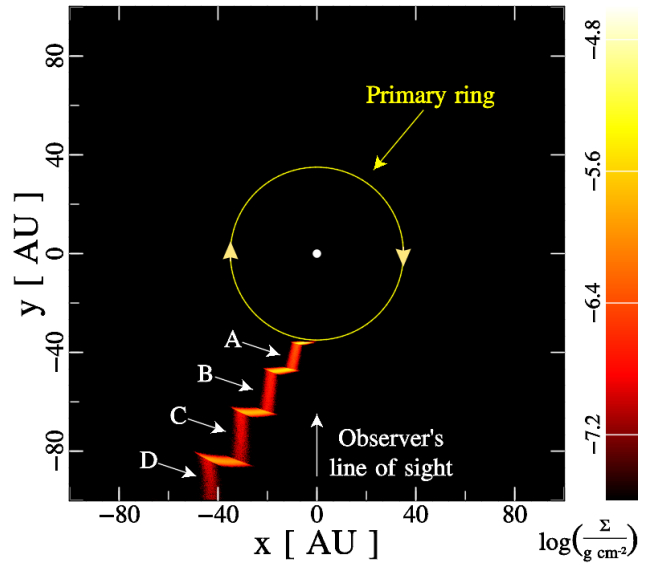}
\caption{Face-on column density $\Sigma$ of simulated
0.1-$\mu$m grains, launched from
the avalanche zone situated on the primary ring, and
accelerated outward by a stellar wind whose mass-loss rate
varies over a 10-yr period. 
The primary ring rotates such that its northwest ansa at $x > 0$
approaches the observer and its southeast ansa at $x < 0$ recedes.
When the wind is weakest
(when $\beta_{\rm w} = \beta_{\rm w,low} = 2$),
grains emanate from the primary ring on more nearly azimuthal
trajectories; these azimuthal segments retain their orientation
as they are blown outward from the star.
The radial segments, emerging from the primary ring
when the wind blows strongest
($\beta_{\rm w} = \beta_{\rm w,high} = 20$) and here labeled A to D,
align with the observer's line of sight;
when viewed edge-on (Figure \ref{fig:edge}),
the radial segments have greater projected column densities
and appear as bright clouds.}
\label{fig:face}
\end{figure}

Figure \ref{fig:face} plots the face-on column density of
particles from our simulation using $\beta_{\rm w,high} = 20$.
The particles trace a zig-zag
path as they flow out of the primary ring.
The azimuthal segments (zigs) correspond to seeds (+ their subsequent
avalanche products) born during the stellar wind's low state:
these particles exited the intersection region moving primarily 
azimuthally. Each initially azimuthal segment was then 
blown outward when the stellar wind later entered
a high state, retaining their azimuthal orientation
for subsequent wind cycles (once a zig,
always a zig). Conversely, radially oriented segments (zags)
contain the seeds (+ their subsequent avalanche products)
born during the stellar wind's high state: these particles
exited the intersection region moving primarily radially, and remain
nearly radial in their orientation as they are blown outward,
aside from a small rotational shear (once a zag, always a zag).

\begin{figure*}[]
\includegraphics[width=1.99\columnwidth]{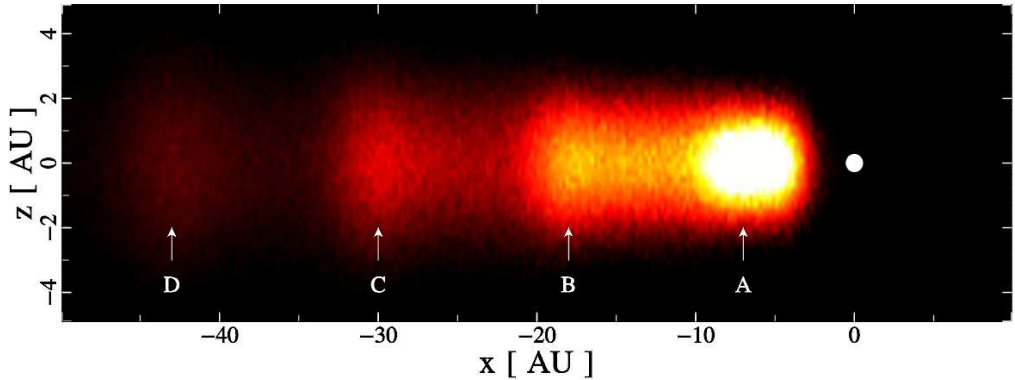}
\caption{Edge-on view of the same outflowing grains as in 
Figure \ref{fig:face}, here showing scattered-starlight surface brightness
instead of raw column density. Cloud A
has a surface brightness that is $\sim$$10\times$ higher than
cloud D. Compare with Figure 2 of \citet{boccaletti15}.}
\label{fig:edge}
\end{figure*}

Remarkably and encouragingly, Figure \ref{fig:edge}
shows that the radial segments appear
as clouds when viewed edge on.
The radial segments have greater line-of-sight column densities
than the azimuthal segments do; thus the radial segments
appear as bright clouds, while the azimuthal segments
represent the inter-cloud regions.
We label our brightest clouds A, B, C and D;
they seem to compare well with features A through D identified
in Figure 2 of \citet{boccaletti15}. In particular,
the projected separations between our simulated clouds
agree with observation.

\begin{figure}[]
\includegraphics[width=0.99\columnwidth]{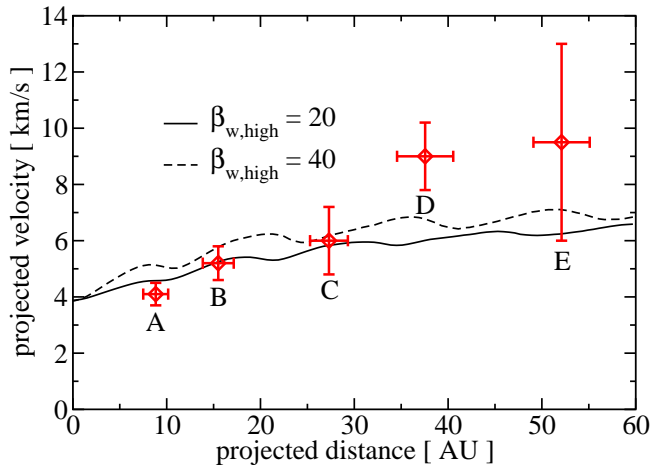}
\caption{Projected velocity profiles of simulated particles,
overplotted with the measured velocities of features A through E
from Figure 4 of \citet{boccaletti15}. Our model agrees well with
observation for the inner three features. Some ideas
for improving the fit for the outer two features
are mentioned in the main text.}
\label{fig:vel}
\end{figure}

Projected velocities of the simulation particles, both for
$\beta_{\rm w,high} = 20$ and 40, are plotted in Figure
\ref{fig:vel}, together with observational data from Figure 4 of
\citet{boccaletti15}. 
The agreement is good for observed features A, B, and C,
and less good for D and E (note that feature E as identified
by \citealt{boccaletti15} is not a bright cloud but an inter-cloud region).
Feature D in particular appears to be something of an outlier,
not much helped by increasing $\beta_{\rm w,high}$.
To improve the fit, we might look to ($i$) decreasing the launch radius
$r_0$ (thereby increasing our model velocities); ($ii$)
incorporating Lorentz forces from the stellar wind
(perhaps we are seeing the onset of magnetic pick-up; see the
last paragraph of \S\ref{pickup});
and ($iii$) independent
re-measurement of feature velocities and re-analysis
of the uncertainties
(see Figure 2 of \citealt{sezestre17} which reports
a new and relatively large error bar on the velocity of feature D.)

\section{SUMMARY AND DISCUSSION}
\label{sum}

We have interpreted the fast-moving features
observed in the AU Mic system as coherent clouds of
dust produced by periodic avalanches.
The clouds are composed of $\sim$0.1-$\mu$m grains,
small enough to experience a radially outward ram pressure force from
the stellar wind up to $\sim$20$\times$ stronger than stellar gravity.
The clouds are launched from a 
region a few AU across---the ``avalanche zone''---situated
on the primary ring encircling the star at $\sim$35 AU, and lying
at an orbital azimuth directly along the line of sight to the star.
The avalanche zone marks where a body of radius $\lesssim$ 400 km and
mass $\lesssim 10^{-4} M_\oplus$ was catastrophically
disrupted less than $\sim$$3 \times 10^4$ yr ago.
The debris from that event, now strewn along a secondary ring
that intersects the primary ring at the
location of the avalanche zone, continues to collide with 
primary ring particles at km/s speeds,
generating the sub-micron grains that seed the avalanche.

This picture can reproduce the individual feature sizes (a few AUs)
and masses ($\sim$$4 \times 10^{-7} M_\oplus$)
inferred from the SPHERE and Hubble Space Telescope
observations, using standard collision parameters 
(e.g., specific energies $\sim$$2 \times 10^8$ erg/g
for catastrophic disruption of competent targets,
and ejecta velocities 
of a few hundred m/s) and ring parameters validated by 
previous modeling of the AU Mic disk (e.g., primary ring
optical depths on the order of $\sim$0.01).
Avalanche amplification factors exceed $\exp(4) \sim 50$.
If avalanches of the kind seen today continue unchecked over
the $\sim$$3\times 10^4$ yr lifetime of the secondary ring,
then a total mass of $\sim$$0.005 M_\oplus$ would be blown
out, roughly half the total mass of the primary ring.
In reality, the avalanches may weaken substantially
well before the secondary ring disintegrates, as the number of
available targets for disruption ($\sim$$\mu$m-sized
parents) in the primary ring drops.
The rate at which mass is lost through avalanches
may eventually asymptote to the rate at which the primary
ring erodes mass through a conventional cascade.

Our theory underscores the potential of dust avalanches
to transfigure debris disks on timescales much shorter
than the age of the star. Dust avalanches leverage the power
of chain reactions and exponential amplification to
process orders of magnitude more mass than they invest.
In our story for AU Mic, avalanches began when a single body less than
a few hundred km in size was shattered.
That single micro-event opened a ``wound'' in the parent
ring that is now hemorrhaging mass on macro-scales.

The stellar wind is expected to be magnetized,
with a field polarity that switches sign according to the
stellar magnetic cycle. The resultant oscillating vertical
Lorentz force (equivalently, the oscillating vertical electric field
seen by grains as the magnetized wind blows past)
will alternately lift and lower cloud grains, which are small enough
to be significantly charged
by stellar ultraviolet photoelectric emission. For plausible
input parameters---ambient field strengths on the order of 10s of
$\mu$G; grain surface potentials on the order of a volt; a magnetic
cycle period of a few years---we can reproduce the
observed magnitude of the vertical displacements. However, the detailed
phasing of the vertical oscillations with time, and their observed
decay with increasing projected separation, require further investigation.

We have not specified with confidence the mechanism regulating
the avalanche period $t_{\rm cycle}$ (equation \ref{eq:tcycle}).
We have supposed that it could set by time variability
of the host stellar wind. Certainly AU Mic's
wind is known to be at least some two orders of magnitude stronger
than the solar wind, as judged by the star's flaring activity, and
from detailed models of the primary ring which point to
significant sculpting by the wind \citep{augereau06, schuppler15}.
But to explain how the dust avalanches turn on and off,
the wind would need to sustain
$\sim$yr-long gusts every $\sim$10 yr, causing the star
to lose mass at peak rates several thousand times
larger than the solar mass loss rate. 
Whether such extreme and sustained
episodes of stellar mass loss are actually realized, and whether
we can reconcile the stellar mass-loss cycle with the stellar
magnetic cycle (as traced by the cloud vertical displacements; see above),
are outstanding issues.

We wonder whether we might dispense with the need for 
decadal-timescale
variability in the stellar mass-loss rate by positing instead
some kind of limit-cycle instability
in the avalanche zone. The order-of-magnitude similarity between
the required cycle period ($\sim$10 yr) 
and the time it takes for 
the avalanche to propagate radially
across the parent ring ($\sim$5 yr from equation \ref{eq:prop})
suggests that perhaps the avalanche zone regulates 
itself---that avalanches
are triggered when some threshold condition 
is periodically satisfied in the intersection region.
It must be some condition on the
seed optical depth. Conceivably the avalanche evacuates the
zone so thoroughly of seeds that the system needs time to re-fill.
Avalanches are characterized by exponential amplification, 
and with exponentials there is extreme sensitivity
to environmental conditions.
At the moment we are unable to say more than this, but the 
possibility
of a self-regulating limit cycle (and a stellar wind 
that is steady but
that still needs to blow strongly to achieve the large
force ratios
$\beta_{\rm w} \sim 10$ 
implied by 
the observed cloud velocities) seems deserving of more thought.
Note that while distinct clouds are created from line-of-sight
projection effects in our time-variable stellar-wind 
model (\S\ref{sim}), they would be created instead by 
time variability in the avalanche dust production rate
in the limit-cycle picture.

Regardless of what drives the time variability,
the avalanche zone from which dust clouds are launched
remains fixed in inertial space. By contrast, orbiting sources of
escaping dust (e.g., a planet---putting aside the separate problem
of how a planet
could be a source of dust in the first place) tend to follow
the dust that
they eject, since the observed velocities of the features 
are comparable to
orbital velocities at $\sim$35 AU. This similarity of velocities leads
to immediate difficulties
in using a moving source to reproduce the observed spacing between
features; the resultant clouds will be too closely spaced
if launched from a moving
source near the primary ring at $\sim$35 AU. Our model avoids
this problem altogether because the dust launch zone is located
at the intersection of two rings; this node does not move, aside from
a negligible precession.

Although our proposal contains significant room for improvement,
it points to a few observational predictions:
($i$) the escaping
cloud grains should have smaller sizes (the better to be accelerated
outward, and to be electrically charged) than their counterparts bound
to the primary and secondary rings; the size difference could be confirmed
by measuring color differences between the fast-moving features
and the rest of the disk
(a pioneering attempt to spatially resolve color differences has been made using the Hubble Space Telescope by \citealt{lomax17});
($ii$) assuming there is only a single avalanche zone
(i.e., a single intersection point between the primary and secondary rings)
located directly along the line of sight to the star,
all escaping clouds will always be seen on the southeast ansa of the disk;
($iii$) the primary ring should rotate such that the northwest ansa
approaches the Earth while the southeast ansa recedes from it.
This sense of rotation is not a crucial detail of our model,
but is preferred because it situates the avalanche zone
on the side of the primary ring nearest the observer,
so that the clouds launched from there are seen more easily
in forward-scattered than in back-scattered starlight.
Spectral line observations in CO gas can test our expectation;
($iv$) the secondary ring has a mass $< 1$\% that of the primary,
and might be distinguished from the primary (this is admittedly ambitious) in ultra-deep exposures
if the rings are mutually inclined; and
($v$) what goes up should come down: the 
features should be observed to vary their vertical
positions according to the stellar magnetic cycle.

\acknowledgments
We thank Pawel Artymowicz, Jean-Charles Augereau, Anthony Boccaletti, 
Rob de Rosa, Tom Esposito, James Graham, Meredith Hughes, 
Paul Kalas, Quentin Kral, Eve Lee, Jamie Lomax, 
Maxwell Millar-Blanchaer, Ruth Murray-Clay, \'Elie Sezestre, 
Kate Su, Philippe Th\'ebault, Jason Wang, and Mark Wyatt 
for discussions.
Erika Nesvold provided a lucid, helpful, and encouraging referee's report.
This work was performed under contract with the Jet Propulsion
Laboratory (JPL) funded by NASA through the Sagan Fellowship Program
executed by the NASA Exoplanet Science Institute.

\bibliography{debris}

\begin{thebibliography}{}
\expandafter\ifx\csname natexlab\endcsname\relax\def\natexlab#1{#1}\fi

\bibitem[{{Artymowicz}(1997)}]{artymowicz97}
{Artymowicz}, P. 1997, Annual Review of Earth and Planetary Sciences, 25, 175

\bibitem[{{Augereau} \& {Beust}(2006)}]{augereau06}
{Augereau}, J.-C., \& {Beust}, H. 2006, \aap, 455, 987

\bibitem[{{Beichman} {et~al.}(2005){Beichman}, {Bryden}, {Gautier},
  {Stapelfeldt}, {Werner}, {Misselt}, {Rieke}, {Stansberry}, \&
  {Trilling}}]{beichman05}
{Beichman}, C.~A., {Bryden}, G., {Gautier}, T.~N., {et~al.} 2005, \apj, 626,
  1061

\bibitem[{{Boccaletti} {et~al.}(2015){Boccaletti}, {Thalmann}, {Lagrange},
  {Janson}, {Augereau}, {Schneider}, {Milli}, {Grady}, {Debes}, {Langlois},
  {Mouillet}, {Henning}, {Dominik}, {Maire}, {Beuzit}, {Carson}, {Dohlen},
  {Engler}, {Feldt}, {Fusco}, {Ginski}, {Girard}, {Hines}, {Kasper}, {Mawet},
  {M{\'e}nard}, {Meyer}, {Moutou}, {Olofsson}, {Rodigas}, {Sauvage},
  {Schlieder}, {Schmid}, {Turatto}, {Udry}, {Vakili}, {Vigan}, {Wahhaj}, \&
  {Wisniewski}}]{boccaletti15}
{Boccaletti}, A., {Thalmann}, C., {Lagrange}, A.-M., {et~al.} 2015, \nat, 526,
  230

\bibitem[{{Brown} {et~al.}(2013){Brown}, {Baliber}, {Bianco}, {Bowman},
  {Burleson}, {Conway}, {Crellin}, {Depagne}, {De Vera}, {Dilday}, {Dragomir},
  {Dubberley}, {Eastman}, {Elphick}, {Falarski}, {Foale}, {Ford}, {Fulton},
  {Garza}, {Gomez}, {Graham}, {Greene}, {Haldeman}, {Hawkins}, {Haworth},
  {Haynes}, {Hidas}, {Hjelstrom}, {Howell}, {Hygelund}, {Lister}, {Lobdill},
  {Martinez}, {Mullins}, {Norbury}, {Parrent}, {Paulson}, {Petry}, {Pickles},
  {Posner}, {Rosing}, {Ross}, {Sand}, {Saunders}, {Shobbrook}, {Shporer},
  {Street}, {Thomas}, {Tsapras}, {Tufts}, {Valenti}, {Vander Horst}, {Walker},
  {White}, \& {Willis}}]{brown13}
{Brown}, T.~M., {Baliber}, N., {Bianco}, F.~B., {et~al.} 2013, \pasp, 125, 1031

\bibitem[{{Cohen}(2011)}]{cohen11}
{Cohen}, O. 2011, \mnras, 417, 2592

\bibitem[{{Dohnanyi}(1969)}]{dohnanyi69}
{Dohnanyi}, J.~S. 1969, \jgr, 74, 2531

\bibitem[{{G{\'a}sp{\'a}r} {et~al.}(2013){G{\'a}sp{\'a}r}, {Rieke}, \&
  {Balog}}]{gaspar13}
{G{\'a}sp{\'a}r}, A., {Rieke}, G.~H., \& {Balog}, Z. 2013, \apj, 768, 25

\bibitem[{{Gopalswamy} {et~al.}(2003){Gopalswamy}, {Lara}, {Yashiro}, {Nunes},
  \& {Howard}}]{gopalswamy03}
{Gopalswamy}, N., {Lara}, A., {Yashiro}, S., {Nunes}, S., \& {Howard}, R.~A.
  2003, in ESA Special Publication, Vol. 535, Solar Variability as an Input to
  the Earth's Environment, ed. A.~{Wilson}, 403--414

\bibitem[{{Graham} {et~al.}(2007){Graham}, {Kalas}, \& {Matthews}}]{graham07}
{Graham}, J.~R., {Kalas}, P.~G., \& {Matthews}, B.~C. 2007, \apj, 654, 595

\bibitem[{{Grigorieva} {et~al.}(2007){Grigorieva}, {Artymowicz}, \&
  {Th{\'e}bault}}]{grigorieva07}
{Grigorieva}, A., {Artymowicz}, P., \& {Th{\'e}bault}, P. 2007, \aap, 461, 537

\bibitem[{{Jackson} {et~al.}(2014){Jackson}, {Wyatt}, {Bonsor}, \&
  {Veras}}]{jackson14}
{Jackson}, A.~P., {Wyatt}, M.~C., {Bonsor}, A., \& {Veras}, D. 2014, \mnras,
  440, 3757

\bibitem[{{Kennedy} \& {Wyatt}(2013)}]{kennedy13}
{Kennedy}, G.~M., \& {Wyatt}, M.~C. 2013, \mnras, 433, 2334

\bibitem[{{Kenyon} \& {Bromley}(2005)}]{kenyon05}
{Kenyon}, S.~J., \& {Bromley}, B.~C. 2005, \aj, 130, 269

\bibitem[{{Kimura} \& {Mann}(1998)}]{kimura98}
{Kimura}, H., \& {Mann}, I. 1998, \apj, 499, 454

\bibitem[{{Kral} {et~al.}(2015){Kral}, {Th{\'e}bault}, {Augereau},
  {Boccaletti}, \& {Charnoz}}]{kral15}
{Kral}, Q., {Th{\'e}bault}, P., {Augereau}, J.-C., {Boccaletti}, A., \&
  {Charnoz}, S. 2015, \aap, 573, A39

\bibitem[{{Krist} {et~al.}(2005){Krist}, {Ardila}, {Golimowski}, {Clampin},
  {Ford}, {Illingworth}, {Hartig}, {Bartko}, {Ben{\'{\i}}tez}, {Blakeslee},
  {Bouwens}, {Bradley}, {Broadhurst}, {Brown}, {Burrows}, {Cheng}, {Cross},
  {Demarco}, {Feldman}, {Franx}, {Goto}, {Gronwall}, {Holden}, {Homeier},
  {Infante}, {Kimble}, {Lesser}, {Martel}, {Mei}, {Menanteau}, {Meurer},
  {Miley}, {Motta}, {Postman}, {Rosati}, {Sirianni}, {Sparks}, {Tran},
  {Tsvetanov}, {White}, \& {Zheng}}]{krist05}
{Krist}, J.~E., {Ardila}, D.~R., {Golimowski}, D.~A., {et~al.} 2005, \aj, 129,
  1008

\bibitem[{{Leinhardt} \& {Stewart}(2012)}]{leinhardt12}
{Leinhardt}, Z.~M., \& {Stewart}, S.~T. 2012, \apj, 745, 79

\bibitem[{{L{\"o}hne} {et~al.}(2008){L{\"o}hne}, {Krivov}, \&
  {Rodmann}}]{lohne08}
{L{\"o}hne}, T., {Krivov}, A.~V., \& {Rodmann}, J. 2008, \apj, 673, 1123

\bibitem[{{Lomax} {et~al.}(2017){Lomax}, {Wisniewski}, {Roberge}, {Donaldson},
  {Debes}, {Malumuth}, \& {Weinberger}}]{lomax17}
{Lomax}, J.~R., {Wisniewski}, J.~P., {Roberge}, A., {et~al.} 2017, ArXiv
  e-prints, arXiv:1705.09291

\bibitem[{{Lykawka} {et~al.}(2012){Lykawka}, {Horner}, {Mukai}, \&
  {Nakamura}}]{lykawka12}
{Lykawka}, P.~S., {Horner}, J., {Mukai}, T., \& {Nakamura}, A.~M. 2012, \mnras,
  421, 1331

\bibitem[{{Mamajek} \& {Bell}(2014)}]{mamajek14}
{Mamajek}, E.~E., \& {Bell}, C.~P.~M. 2014, \mnras, 445, 2169

\bibitem[{{Matthews} {et~al.}(2014){Matthews}, {Krivov}, {Wyatt}, {Bryden}, \&
  {Eiroa}}]{matthews14}
{Matthews}, B.~C., {Krivov}, A.~V., {Wyatt}, M.~C., {Bryden}, G., \& {Eiroa},
  C. 2014, Protostars and Planets VI, 521

\bibitem[{{Matthews} {et~al.}(2015){Matthews}, {Kennedy}, {Sibthorpe},
  {Holland}, {Booth}, {Kalas}, {MacGregor}, {Wilner}, {Vandenbussche},
  {Olofsson}, {Blommaert}, {Brandeker}, {Dent}, {de Vries}, {Di Francesco},
  {Fridlund}, {Graham}, {Greaves}, {Heras}, {Hogerheijde}, {Ivison}, {Pantin},
  \& {Pilbratt}}]{matthews15}
{Matthews}, B.~C., {Kennedy}, G., {Sibthorpe}, B., {et~al.} 2015, \apj, 811,
  100

\bibitem[{{Melis} {et~al.}(2012){Melis}, {Zuckerman}, {Rhee}, {Song}, {Murphy},
  \& {Bessell}}]{melis12}
{Melis}, C., {Zuckerman}, B., {Rhee}, J.~H., {et~al.} 2012, \nat, 487, 74

\bibitem[{{Meng} {et~al.}(2014){Meng}, {Su}, {Rieke}, {Stevenson}, {Plavchan},
  {Rujopakarn}, {Lisse}, {Poshyachinda}, \& {Reichart}}]{meng14}
{Meng}, H.~Y.~A., {Su}, K.~Y.~L., {Rieke}, G.~H., {et~al.} 2014, Science, 345,
  1032

\bibitem[{{Pan} \& {Sari}(2005)}]{pan05}
{Pan}, M., \& {Sari}, R. 2005, \icarus, 173, 342

\bibitem[{{Reinhold} {et~al.}(2017){Reinhold}, {Cameron}, \&
  {Gizon}}]{reinhold17}
{Reinhold}, T., {Cameron}, R.~H., \& {Gizon}, L. 2017, \aap, 603, A52

\bibitem[{{Robinson} {et~al.}(2001){Robinson}, {Linsky}, {Woodgate}, \&
  {Timothy}}]{robinson01}
{Robinson}, R.~D., {Linsky}, J.~L., {Woodgate}, B.~E., \& {Timothy}, J.~G.
  2001, \apj, 554, 368

\bibitem[{{Schneider} {et~al.}(2014){Schneider}, {Grady}, {Hines}, {Stark},
  {Debes}, {Carson}, {Kuchner}, {Perrin}, {Weinberger}, {Wisniewski},
  {Silverstone}, {Jang-Condell}, {Henning}, {Woodgate}, {Serabyn},
  {Moro-Martin}, {Tamura}, {Hinz}, \& {Rodigas}}]{schneider14}
{Schneider}, G., {Grady}, C.~A., {Hines}, D.~C., {et~al.} 2014, \aj, 148, 59

\bibitem[{{Sch{\"u}ppler} {et~al.}(2015){Sch{\"u}ppler}, {L{\"o}hne}, {Krivov},
  {Ertel}, {Marshall}, {Wolf}, {Wyatt}, {Augereau}, \& {Metchev}}]{schuppler15}
{Sch{\"u}ppler}, C., {L{\"o}hne}, T., {Krivov}, A.~V., {et~al.} 2015, \aap,
  581, A97

\bibitem[{{Schwenn}(2000)}]{schwenn00}
{Schwenn}, R. 2000, {Solar Wind: Global Properties}, ed. P.~{Murdin}

\bibitem[{{Sezestre} {et~al.}(2017){Sezestre}, {Augereau}, {Boccaletti}, \&
  {Th{\'e}bault}}]{sezestre17}
{Sezestre}, {\'E}., {Augereau}, J.-C., {Boccaletti}, A., \& {Th{\'e}bault}, P.
  2017, ArXiv e-prints, arXiv:1707.09761

\bibitem[{{Shen} {et~al.}(2009){Shen}, {Draine}, \& {Johnson}}]{shen09}
{Shen}, Y., {Draine}, B.~T., \& {Johnson}, E.~T. 2009, \apj, 696, 2126

\bibitem[{{Song} {et~al.}(2005){Song}, {Zuckerman}, {Weinberger}, \&
  {Becklin}}]{song05}
{Song}, I., {Zuckerman}, B., {Weinberger}, A.~J., \& {Becklin}, E.~E. 2005,
  \nat, 436, 363

\bibitem[{{Strubbe} \& {Chiang}(2006)}]{strubbe06}
{Strubbe}, L.~E., \& {Chiang}, E.~I. 2006, \apj, 648, 652

\bibitem[{{Takasawa} {et~al.}(2011){Takasawa}, {Nakamura}, {Kadono}, {Arakawa},
  {Dohi}, {Ohno}, {Seto}, {Maeda}, {Shigemori}, {Hironaka}, {Sakaiya},
  {Fujioka}, {Sano}, {Otani}, {Watari}, {Sangen}, {Setoh}, {Machii}, \&
  {Takeuchi}}]{takasawa11}
{Takasawa}, S., {Nakamura}, A.~M., {Kadono}, T., {et~al.} 2011, \apjl, 733, L39

\bibitem[{{Tielens} {et~al.}(1994){Tielens}, {McKee}, {Seab}, \&
  {Hollenbach}}]{tielens94}
{Tielens}, A.~G.~G.~M., {McKee}, C.~F., {Seab}, C.~G., \& {Hollenbach}, D.~J.
  1994, \apj, 431, 321

\bibitem[{{Wang} {et~al.}(2015){Wang}, {Graham}, {Pueyo}, {Nielsen},
  {Millar-Blanchaer}, {De Rosa}, {Kalas}, {Ammons}, {Bulger}, {Cardwell},
  {Chen}, {Chiang}, {Chilcote}, {Doyon}, {Draper}, {Duch{\^e}ne}, {Esposito},
  {Fitzgerald}, {Goodsell}, {Greenbaum}, {Hartung}, {Hibon}, {Hinkley}, {Hung},
  {Ingraham}, {Larkin}, {Macintosh}, {Maire}, {Marchis}, {Marois}, {Matthews},
  {Morzinski}, {Oppenheimer}, {Patience}, {Perrin}, {Rajan}, {Rantakyr{\"o}},
  {Sadakuni}, {Serio}, {Sivaramakrishnan}, {Soummer}, {Thomas}, {Ward-Duong},
  {Wiktorowicz}, \& {Wolff}}]{wang15}
{Wang}, J.~J., {Graham}, J.~R., {Pueyo}, L., {et~al.} 2015, \apjl, 811, L19

\bibitem[{{Weber} \& {Davis}(1967)}]{weber67}
{Weber}, E.~J., \& {Davis}, Jr., L. 1967, \apj, 148, 217

\bibitem[{{Weinberger} {et~al.}(2011){Weinberger}, {Becklin}, {Song}, \&
  {Zuckerman}}]{weinberger11}
{Weinberger}, A.~J., {Becklin}, E.~E., {Song}, I., \& {Zuckerman}, B. 2011,
  \apj, 726, 72

\bibitem[{{Wyatt}(2008)}]{wyatt08}
{Wyatt}, M.~C. 2008, \araa, 46, 339

\bibitem[{{Wyatt} {et~al.}(2011){Wyatt}, {Clarke}, \& {Booth}}]{wyatt11}
{Wyatt}, M.~C., {Clarke}, C.~J., \& {Booth}, M. 2011, Celestial Mechanics and
  Dynamical Astronomy, 111, 1

\bibitem[{{Wyatt} {et~al.}(2005){Wyatt}, {Greaves}, {Dent}, \&
  {Coulson}}]{wyatt05}
{Wyatt}, M.~C., {Greaves}, J.~S., {Dent}, W.~R.~F., \& {Coulson}, I.~M. 2005,
  \apj, 620, 492

\end{thebibliography}
\end{document}